# Effects of calcium substitution on the superconducting properties of $R_{1-x}Ca_xBa_2Cu_3O_z$ ($R$ = Eu, Gd, Er; $0 \le x \le 0.3$) polycrystalline samples

E. K. Nazarova [a,*C], K. Nenkov [b,c], G. Fuchs [b], K. - H. Müller [b]

[a] *Institute of Solid State Physics, Bulgarian Academy of Sciences, 72 Tzarigradsko Chaussee Blvd. 1784 Sofia , Bulgaria*

[b] *IFW, Leibniz Institute for Solid State and Materials Research Dresden, P.O. BOX 270016, D-01171 Dresden, Germany*

[c] *International Laboratory of High Magnetic Fields and Low Temperatures, Gajowicka 95, 53-529 Wroclaw, Poland*

## Abstract

The influence of calcium substitution on the superconducting properties of polycrystalline $R_{1-x}Ca_xBa_2Cu_3O_z$ ($R$ = Eu, Gd, Er; x=0; 0.2; 0.25 and 0.3) samples has been studied by X-ray powder diffraction, *ac* susceptibility and *dc* magnetization measurements. The superconducting parameters such as critical temperature, inter- and intra-granular critical current and flux pinning are found to be strongly dependent both on Ca content and type of $R$ element. The best combination of these parameters is found for the system $Gd_{1-x}Ca_xBa_2Cu_3O_z$ forming $R_{1+y}Ba_{2-y}Cu_3O_z$ clusters. The level of overdoping and the type of intergrain connection, were found to be influenced by the $R$ element and the Ca concentration. Flux pinning in $Gd_{1-x}Ca_xBa_2Cu_3O_z$ is connected with the presence of $R_{1+y}Ba_{2-y}Cu_3O_z$ clusters.

---

[C] orresponding author. Tel.: +359-2-7144-211; Fax: +359-2-975-36-32.
E-mail address: nazarova@issp.bas.bg (E. K. Nazarova)

## Introduction

The recent achievements of critical current density exceeding $10^6$ A/cm$^2$ at 77 K in YBCO coated conductors have stimulated interest in their practical applications at high temperature and high magnetic field [1-4]. The main factors reducing the $J_c$ in HTSC are grain boundaries (GB) and poor flux pinning [5]. One concept for solving the GB problem was to fabricate materials such as the single-crystal-like epitaxial thin films, melt-textured monoliths, ect. in order to reduce the number of GBs. This concept is applicable for short samples preparation. However, it is difficult to produce long tapes and wires without relatively high angle GBs suppressing the $J_c$ significantly [6]. An alternative in this case is the improvement of grain boundary connections by chemical doping [7]. The bicrystal technology was successfully applied for investigation of interface's properties. A. Schmehl et al. [8] reported recently that by doping $Ca^{2+}$ at $Y^{3+}$ site, the weak link effect of high angle GBs in YBCO multi-layer films was significantly reduced and the critical current density was enhanced more than 7 times. The GBs presenting in superconducting layers of coated conductors behave in many ways like bicrystal GBs [9] and Ca doping was applied for increasing $J_c$ of coated conductors [10]. Improvement of grain boundary transport in bulk melt-processed [11] and sintered [12] YBCO samples with Ca substitution have been also established. Although the effect is smaller than in thin films there are some indications for improved flux pinning.

In this work we report our investigations concerning Ca substitution in $R$Ba$_2$Cu$_3$O$_z$ ($R$ = Eu, Gd and Er) systems from the viewpoint of carrier concentration and flux pinning.

## 2. Experimental

*Sample preparation*: Three series of $R_{1-x}$Ca$_x$Ba$_2$Cu$_3$O$_z$ samples ($R$ = Eu, Gd and Er) were prepared. Each series consists of four different compositions with x=0; 0.2; 0.25 and 0.3. The standard solid-state reaction method is used for samples preparation. Grinding and heating steps were repeated three times. The first heating was at 900 °C for 21 h in flowing oxygen. The second one was at 930 °C at the same atmosphere followed by slow cooling and additional annealing at 450 °C for 2 h. Tablets were pressed, heated up to 950 °C for 23 h and subsequently annealed at 450 °C for 23 h. The density of synthesized samples was found to vary between 70% and 90% of the corresponding theoretical value and depends on the kind of Re element and Ca substitution. Ca free samples have larger grains and are more porous than substituted samples.

*Investigations*: Standard X-ray powder diffraction analysis with $Co_{k\alpha}$ radiation was used for the examination of material structural properties at room temperature.

AC magnetic susceptibility was measured with commercial Lake Shore-7000 Susceptometer in a field of 0.1 Oe at 133 Hz. DC magnetization was investigated using a Quantum Design SQUID Magnetometer. The full hysteresis cycle was recorded after the specimen is zero field cooled to the desired temperature.

Temperature dependence of critical current density at zero magnetic field was measured in Physical Properties Measurement System – Quantum Design. For these measurements samples were cut into bars. Their dimensions and distance between the current leads were accurately measured. 1μV/cm criterion was used to determine the critical current.

## 3. Results and discussion

### 3.1.Carrier concentration

Figure 1 shows room temperature X-ray diffraction patterns for $Eu_{1-x}Ca_xBa_2Cu_3O_z$ (with x=0; 0.2; 0.25 and 0.3) samples. Identical pictures (not shown here) have been obtained for the compounds with *R*=Gd and Er. Independently of the R element, all samples with x=0 and x=0.2 are single-phase samples of 123 orthorhombic structure. By detailed analysis in the range of $2\theta \approx 30-40°$, detectable amounts of $BaCuO_2$ were found in samples with a Ca content of $x \geq 0.25$. The appearance of the $BaCuO_2$ impurity phase indicates that Ca starts to substitute not only for the *R* element but also for Ba [13-14]. Thus some cation deficiency on *R* and Cu positions will take place.

With increasing **x** the double peaks at 2θ ≈ 54.5° show the tendency towards a triplet peak structure (Fig. 1), which has been identified with the appearance of the ortho-II phase [15-16].

The obtained data for the unit cell lattice parameters and the orthorhombic splitting (*b*-*a*) are listed in Table 1. It has been found that, independently of the type of R species, the lattice parameter *a* remains almost stable. Consistently with the appearance of the ortho-II phase, the lattice parameter *b* and the orthorhombic splitting decrease with increasing Ca concentration. This means that the structure losses oxygen from the chains and the total oxygen content decreases monotonously with increasing **x**. This is also supported by the increasing value of the lattice parameter *c* [16] and has been established by oxygen contents measurements for different $R_{1-x}Ca_xBa_2Cu_3O_z$ compounds [12,17-18].

In Fig.2, data of AC susceptibility are presented. The onset of the superconducting transition has been used to determine $T_{c\ onset}$ (see Table 1). It is well known that $T_c$ of high temperature superconductors follows an approximately parabolic dependence upon the doped holes concentration, *p* [19]. For optimally doped $RBa_2Cu_3O_z$ compounds the critical temperature is found to be maximal ($T_{c\ max}$) at $p$≈0.16 holes per $CuO_2$ planes. Using the experimental data from [20] we estimated $T_{c\ max}$ to be 94 K for *R*=Eu and Gd and $T_{c\ max}$ = 92 K for *R*=Er. For Ca substituted samples with x=0.2 we estimated $T_{c\ max}$=87.0 K for *R*=Eu; $T_{cmax}$=85.5 K for *R*=Gd and $T_{c\ max}$=84.7 K for *R*=Er. These values for $T_{c\ max}$ have been marked in Table 1 with ( ) in order to indicate that they have been taken from [20]. From the above cited data we found that the critical temperature decreases in Ca substituted samples with a rate of 0.350 K per at % $Ca^{2+}$ substituting $Eu^{3+}$; with 0.425 K per at % $Ca^{2+}$ substituting $Gd^{3+}$ and with 0.365 K per at % $Ca^{2+}$ substituting $Er^{3+}$. These results are used to calculate the values of $T_{c\ max}$ for Ca substituted samples with x=0.25 and x=0.3. The obtained values are marked in Table 1 with (*). Our non-substituted samples with R=Eu and Er have values of $T_{c\ onset}$ close to $T_{c\ max}$. Their oxygen concentration, determined by spectrophotometric method [21], is in the range of 6.94 - 6.98. It has been shown by Tallon at all [19] that such samples are slightly overdoped. The largest deviation from $T_{c\ max}$ (about 2.3 K) is observed for the Gd123 sample, which has also the largest transition width. This can be attributed to the formation of $R$-riched $R_{1+y}Ba_{2-y}Cu_3O_z$ clusters during the synthesis in oxygen atmosphere [22,23]. Due to the $R$-Ba substitution, these superconductors usually exhibit lower $T_c$, which can be explained in terms of a decrease in the carrier concentration when $R$ ions with $3^+$ valence substitutes for Ba with $2^+$ valence. Even when the $T_{c\ onset}$ is high, a broad transition is observed in the AC susceptibility vs. temperature dependence as shown in Fig. 2 b) for Gd123. It is reasonably to assume that the $T_c$ suppression in Gd123 is rather a manifestation of underdoping than of overdoping.

We determine that the oxygen content of samples with a Ca content of x = 0.2 varies between 6.79 for $Gd_{0.8}Ca_{0.2}Ba_2Cu_3O_z$ and 6.86 for $Eu_{0.8}Ca_{0.2}Ba_2Cu_3O_z$. $T_{c\ onset}$ for both specimens is about 80 K. Non substituted Gd123 and Eu123, with approximately the same oxygen content and $T_c$ [20] are underdoped. Therefore substituted samples, due to the additional carriers given by Ca, have to be overdoped and we place their $T_c$ to the right of the maximum of the $T_c(p)$ dependence.

Ca substitution for $R$ = Er requires careful consideration. Er in eight-fold coordination has a smaller ionic radius (IR) of IR=1.004 A than eight-fold coordinated Ca with IR=1.12 A. Therefore, Ca probably substitutes for Er in six-fold coordination with IR=1.0 A which matches better that of the eight-fold coordinated Er. For comparison, the IR in eight-fold coordinated Eu (1.066 A) and Gd (1.053A) are close to that of eightfold coordinated Ca. According to [17] oxygen vacancies are created in $CuO_2$ planes by the substitution of trivalent Er in eight-fold coordination for divalent Ca in six-fold coordination, thus seriously affecting the critical temperature. The question whether the oxygen vacancies are formed in $CuO_2$ planes or in Cu-O chains is controversially discussed [17, 24-25]. It has been shown by neutron diffraction analysis on $Er_{1-x}Ca_xBa_2Cu_3O_y$ ($0 \leq x \leq 0.3$) that the lowering of the overall oxygen content with **x** is not entirely from Cu-O chains, but also partly from the $CuO_2$ planes. From our XRD measurements we can observe only the *(b-a)* suppression due to the oxygen loss from the chains. We find also, by contrast with the other compounds, that a broad intra-granular transition occurs in $Er_{1-x}Ca_xBa_2Cu_3O_y$ samples. In the Fig.2c (inset) details of superconducting transition for $Er_{0.8}Ca_{0.2}Ba_2Cu_3O_z$ are shown. A broad intragranular transition is an indication for the existence of superconducting grains having carrier concentration within large range. This might be a result of oxygen loss from both: chains and planes. In order to verify that this sample is overdoped it has been oxygenated additionally for another 50 h at 450°C in flowing oxygen and we found a drop of the critical temperature to 74.1 K. Similar results were obtained for other $Er_{1-x}Ca_xBa_2Cu_3O_z$ samples with x=0.25 and x=0.3 after such additional oxygenation.

Let's notice also the narrowing of the superconducting transition for sample $Gd_{0.8}Ca_{0.2}Ba_2Cu_3O_z$ compared to that of the Gd123 specimen. A replacement of 20 at % of Gd atoms by Ca was found to compensate for missing carriers due to the *R*-riched clusters formation in Gd123. Thus the $Gd_{0.8}Ca_{0.2}Ba_2Cu_3O_z$ sample has a smaller transition width than. Gd123. Further increasing of Ca substitution (x≥0.25) leads to the formation of non-superconducting $BaCuO_2$ due to the Ca occupation not only on the *R* site but on the Ba site, too. For the samples within the solubility limit of Ca, the hole concentration is determined by the balance of two opposite factors. The carrier concentration increases due to the $R^{3+}$ - $Ca^{2+}$ substitution and decreases because of oxygen loss and $R^{3+}$-$Ba^{2+}$ substitution. Another mechanism for oxygen reduction is possible when a *R* element with smaller IR in eight-fold coordination is substituted by Ca in six-fold coordination. For samples where the limit of Ca solubility is exceeded all these mechanisms are valid, but the appearance of cation deficiency gives an additional mechanism for increasing the holes concentration. As a result the overdoping goes on for all samples with x=0.25 and x=0.3.

We used the well-known empirical expression $T_c/T_{c\,max}=1-82.6(p-0.16)^2$ [19] to determine the hole concentration $p$ per $CuO_2$ planes. It has been shown that irrespective of substitution, $T_c/T_{c\,max}$ remains the same function of hole concentration. The last is independently determined by bond valence sum method [26]. The results obtained for our samples are listed in Table 1 and presented in Fig. 3. Two important conclusions can be drawn from Fig. 3. Firstly, additional holes provided by Ca compensate the carriers` loss due to the formation of *R*-riched clusters in the Gd system. $Gd_{1-x}Ca_xBa_2Cu_3O_z$ samples are slightly overdoped with relatively high critical temperature and $Gd_{1+y}Ba_{2-y}Cu_3O_z$ clusters known as good pinning centers. Secondly, in the frame of each *R* series, samples with x=0.3 are less overdoped than those with x=0.2 and x=0.25 which can be attributed to the preparation conditions.

We measured the current-voltage characteristics at different temperatures and determined the intergranular critical current, $J_{c\ inter}$, as a function of temperature for $Gd_{1-x}Ca_xBa_2Cu_3O_y$ compounds (see Fig. 4). In ceramic materials the critical current is limited by Josephson weak links within the grain boundaries. The $J_c(T)$ dependence can be used to determine the type of the Josephson junction. Superconductor– insulator–superconductor (S-I-S) junctions are formed when $J_c$ is limited by insulating GBs. In this case, the temperature dependence of supercurrent is described by the Ambegaokar–Baratoff formula [27] predicting a linear behavior close to $T_c$ [28]. Fig. 4 shows that samples with x=0 and x=0.2 have a linear $J_c(T)$ dependence.

When $J_c$ is limited by superconductor–normal metal–superconductor (S-N-S) junctions, the maximum Josephson current is given by de Gennes proximity-effect theory [29]. It predicts a quadratic temperature dependence of supercurrents close to $T_c$. We observe a similar $J_c(T)$ dependence for the sample with x=0.3.

Probably some “metallization” of GB regions takes place due to the increasing of Ca concentration. The mechanism of Ca doping at GBs is not clear although some explanations are proposed [7, 30]. The junction resistance is reduced and current goes preferentially through S-N-S junctions. The increasing amount of insulating $BaCuO_2$ phase in this sample shrink the cross section of the current path and, therefore, it shows the smallest intergranular critical current in spite of being overdoped. This result shows, that Ca substitution is able to change not only the carrier concentration, but also the intergrain connections.

### 3.2. Flux pinning

Flux pinning is a very important characteristic for high field applications of high temperature superconductors. Chemical fluctuations are able to create imperfect regions in the *R*-123 matrix with depressed $T_c$ and dimensions of the order of the coherence length. When the applied field is increased these regions are driven normal, since their critical field is lower than that of the matrix. Thus they start to act as effective pinning centers causing the so-called “peak effect” [31].

In order to investigate the influence of Ca substitution on the pinning mechanisms we recorded the magnetization hysteresis loops for all samples after zero field cooling at T=20 K. Hysteresis curves for some samples are presented in Fig. 5. All compounds are superconducting at 20 K and when the field increases the magnetization is negative. It becomes positive, however, at fields about 0.4÷0.6 T for $GdBa_2Cu_3O_z$ and $ErBa_2Cu_3O_z$. For $EuBa_2Cu_3O_z$, the magnetization was found to change its sign at a significantly higher field of about 6.8 T.

It is well known that the magnetic $R^{3+}$ ions in Gd123 and Er123 compounds order antiferromagnetically at 2.2 K [32] and at 0.3 K [33], respectively. Above these temperatures the magnetization is influenced by the paramagnetism of these compounds [34]. From Fig.5 (a) and (c) it is clearly seen that Ca substitution significantly reduces the width of the hysteresis loops of $Eu_{1-x}Ca_xBa_2Cu_3O_z$ and $Er_{1-x}Ca_xBa_2Cu_3O_z$ compounds. Chemical doping in these compounds improves the GB’s $J_c$, but at the same time it strongly degrades the intragrain critical current density $J_{c\ intra}$. Only in $Gd_{1-x}Ca_xBa_2Cu_3O_z$ samples the effect of Ca substitution on $J_{c\ intra}$ remains moderate.

According to the Bean critical state model [35], $J_{c\ intra}$ can be determined from the width of the hysteresis loop [36]: $J_c = 15\ \Delta M/d$, where $\Delta M$ is the difference of the magnetic moments between ascending and descending field branches of the hysteresis in emu/cm$^3$ and $d$ is the average grain size within the ceramic samples in cm. From this expression, $J_{c\ intra}$ is obtained in A/cm$^2$. The grain size of the investigated $Gd_{1-x}Ca_xBa_2Cu_3O_y$ samples estimated from SEM micrographs was found to be in the range of 10 - 14 µm for x=0 and of 8-10 µm for x=0.25. To determine $J_{c\ intra}$ of these samples, an average grain size of 10 µm was used. The field dependence of $J_{c\ intra}$ of the $Gd_{1-x}Ca_xBa_2Cu_3O_y$ samples at T=20 K is presented in Fig. 6. We make a correction for Gd paramagnetism [34] but for clarity of presentation both results for $J_{c\ intra}$ (with and without of paramagnetic correction) have been presented only for Gd123 sample. For the other samples only corrected values of $J_{c\ intra}$ are presented. In Ca substituted samples the value of correction decreases. Correction for paramagnetizm of $BaCuO_2$ phase is reasonably neglected [37].

In order to separate different flux pinning mechanisms in substituted and non-substituted samples, the normalized loop width $M/M_{max}$ was plotted against the reduced field $b= B/B_{max}$ with $B_{max}$ as the field of the maximum loop width $M_{max}$ (see Fig. 7). From this figure we see that all substituted samples are scaled on a single curve different from the curve for the non-substituted sample. This means that pinning in the Ca-substituted specimens has a common nature. Probably, pinning is caused by a combination of *R*-riched clusters [38], Ca substituted nano-sized regions with reduced $T_c$ and oxygen vacancies, which form non-superconducting regions. In general, pinning provided by chemical fluctuations is stronger than that provided by oxygen vacancies [39-40].

## 4. Conclusions

In conclusion we demonstrate that Ca substitution is useful to prepare overdoped 123 samples with *R*=Eu, Gd and Er. It is less effective in systems which form $R_{1+y}Ba_{2-y}Cu_3O_z$ clusters as the holes provided by Ca are partially compensated by the raised number of $R^{3+}$ ions. Thus, a slightly overdoped sample with relatively high $T_c$ and effective pinning centers can be produced. When Ca substitutes a *R* element with large IR the superconducting transition is less influenced than that of systems with small IR of the *R* element.

By the introduction of Ca in $GdBa_2Cu_3O_z$ relatively high intragranular critical currents can be saved. The combination of $Gd_{1+y}Ba_{2-y}Cu_3O_z$ clusters, Ca substituted clusters of reduced $T_c$ and oxygen vacancies result in effective flux pinning in $Gd_{1-x}Ca_xBa_2Cu_3O_z$ compounds. All of the above mentioned effects are expected to be more pronounced in coated conductors where additionally high transport currents can be achieved by the good grain alignment.

## Acknowledgement

The authors are grateful to A. Stoyanova-Ivanova, M. Terzieva and K. Momchilova for samples preparation and oxygen content determination. One of the authors (E. N.) expresses his gratitude to DFG, Germany for the financial support in conducting this work.

## Figure Captions

**Fig.1** X-ray diffraction patterns for $Eu_{1-x}Ca_xBa_2Cu_3O_z$ (with x=0; 0.2; 0.25 and 0.3) samples.

**Fig.2** Temperature dependence of the real part of fundamental AC magnetic susceptibility for all investigated samples (a) $Eu_{1-x}Ca_xBa_2Cu_3O_z$, (b) $Gd_{1-x}Ca_xBa_2Cu_3O_z$ and (c) $Er_{1-x}Ca_xBa_2Cu_3O_z$ at $H_{ac}$= 0.1 Oe and f=133 Hz.

**Fig. 3** Critical temperature, $T_c$ as a function of hole concentration $p$ for all investigated samples.

**Fig. 4** $J_{c,inter}$ vs. $T/T_{c,onset}$ for $Gd_{1-x}Ca_xBa_2Cu_3O_z$ samples. The lines present the best linear fit to the data for samples with x=0 and x=0.2 and the best quadratic fit for the sample with x=0.3.

**Fig. 5** Magnetization hysteresis loops for (a) $Eu_{1-x}Ca_xBa_2Cu_3O_z$, (b) $Gd_{1-x}Ca_xBa_2Cu_3O_z$ and (c) $Er_{1-x}Ca_xBa_2Cu_3O_z$. samples with x=0, 0.2, 0.25 and 0.3. All loops were recorded at T=20 K.

**Fig. 6** Critical current density vs. applied field for $Gd_{1-x}Ca_xBa_2Cu_3O_z$ samples with x=0, 0.2, 0.25 and 0.3 at T=20 K.

**Fig.7** Plots of $M/M_{max}$ versus $B/B_{max}$ for $Gd_{1-x}Ca_xBa_2Cu_3O_z$ samples with x=0, 0.2, 0.25 and 0.3 at T=20 K.

Fig.1.

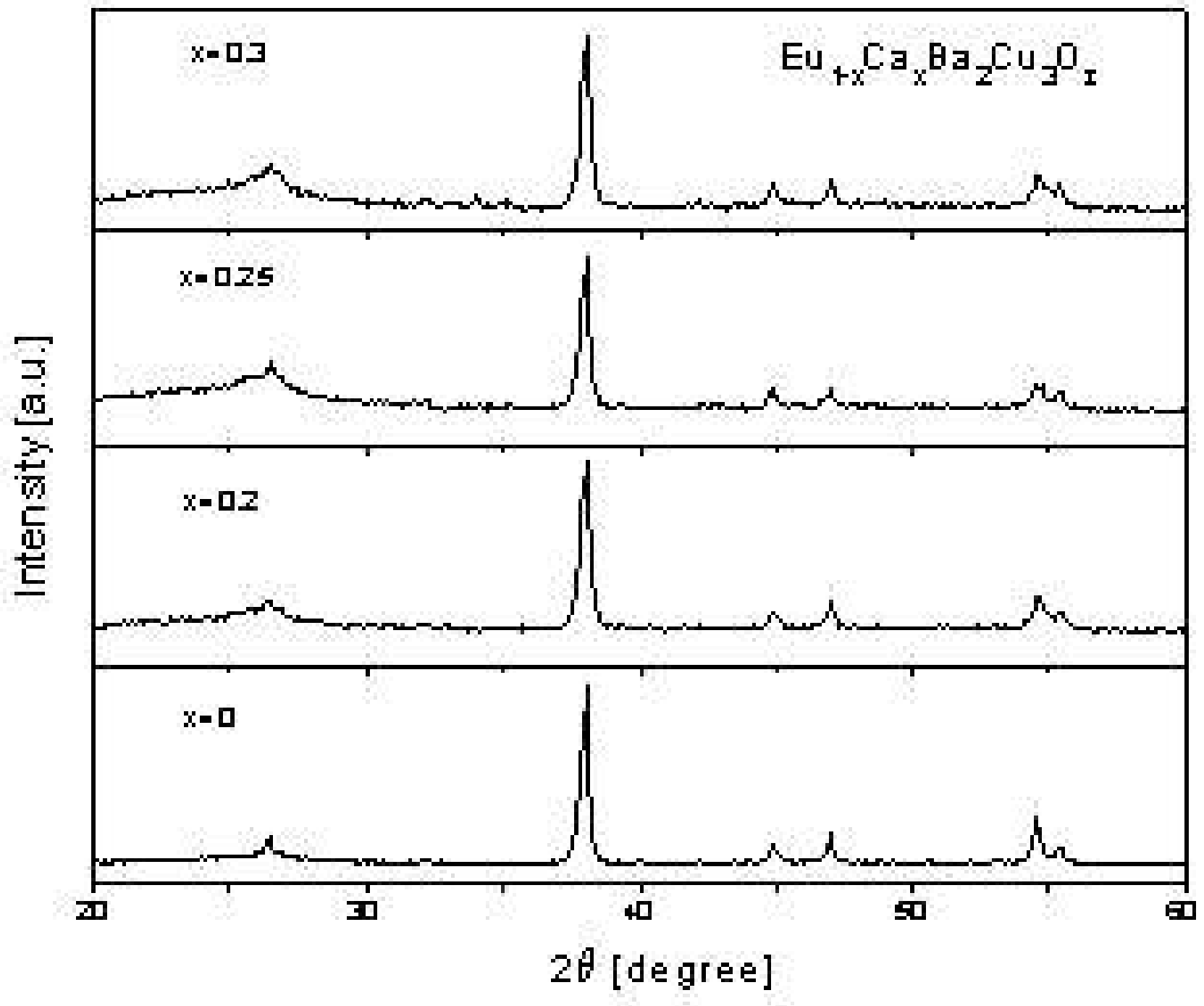

Fig.2a.

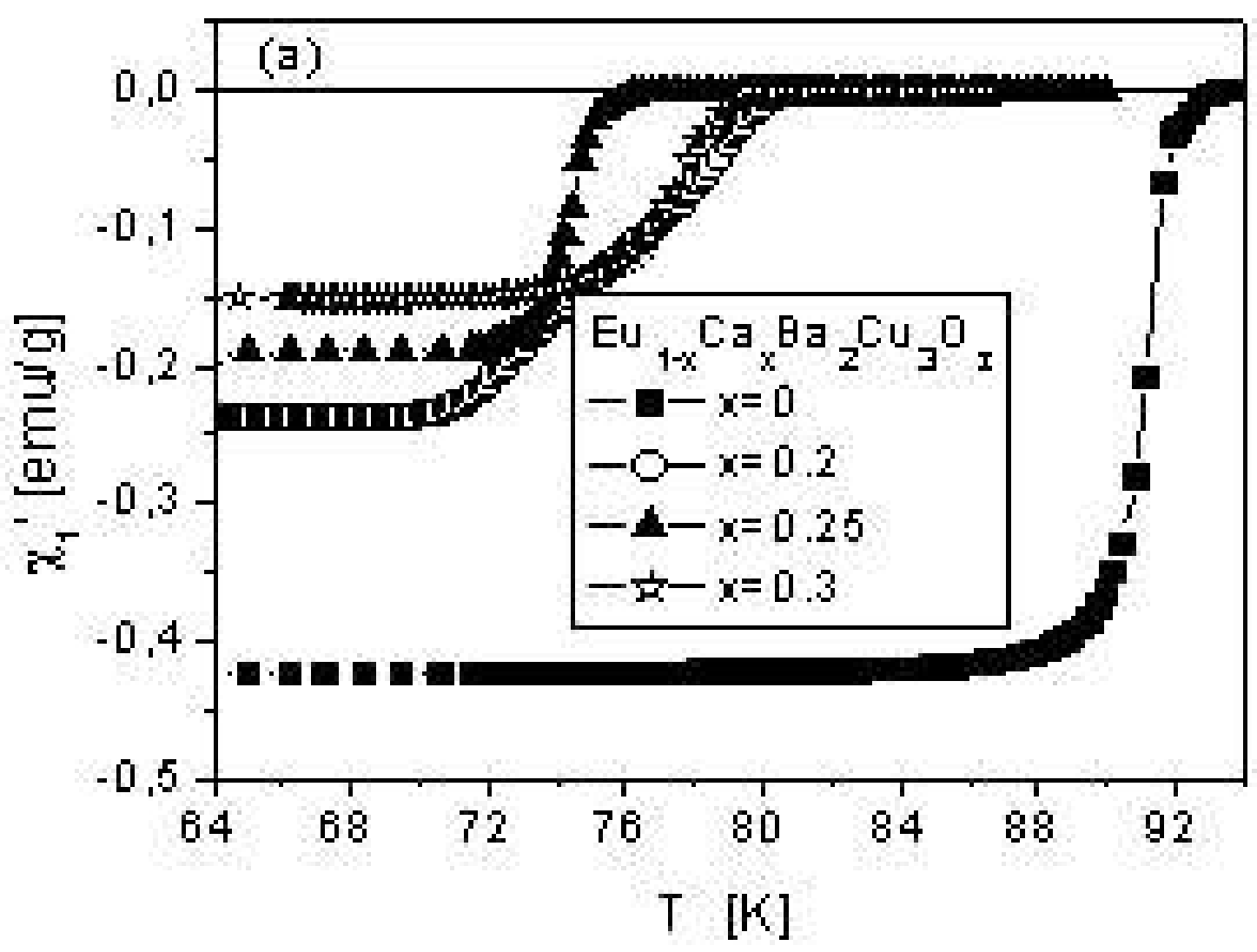

Fig.2b.

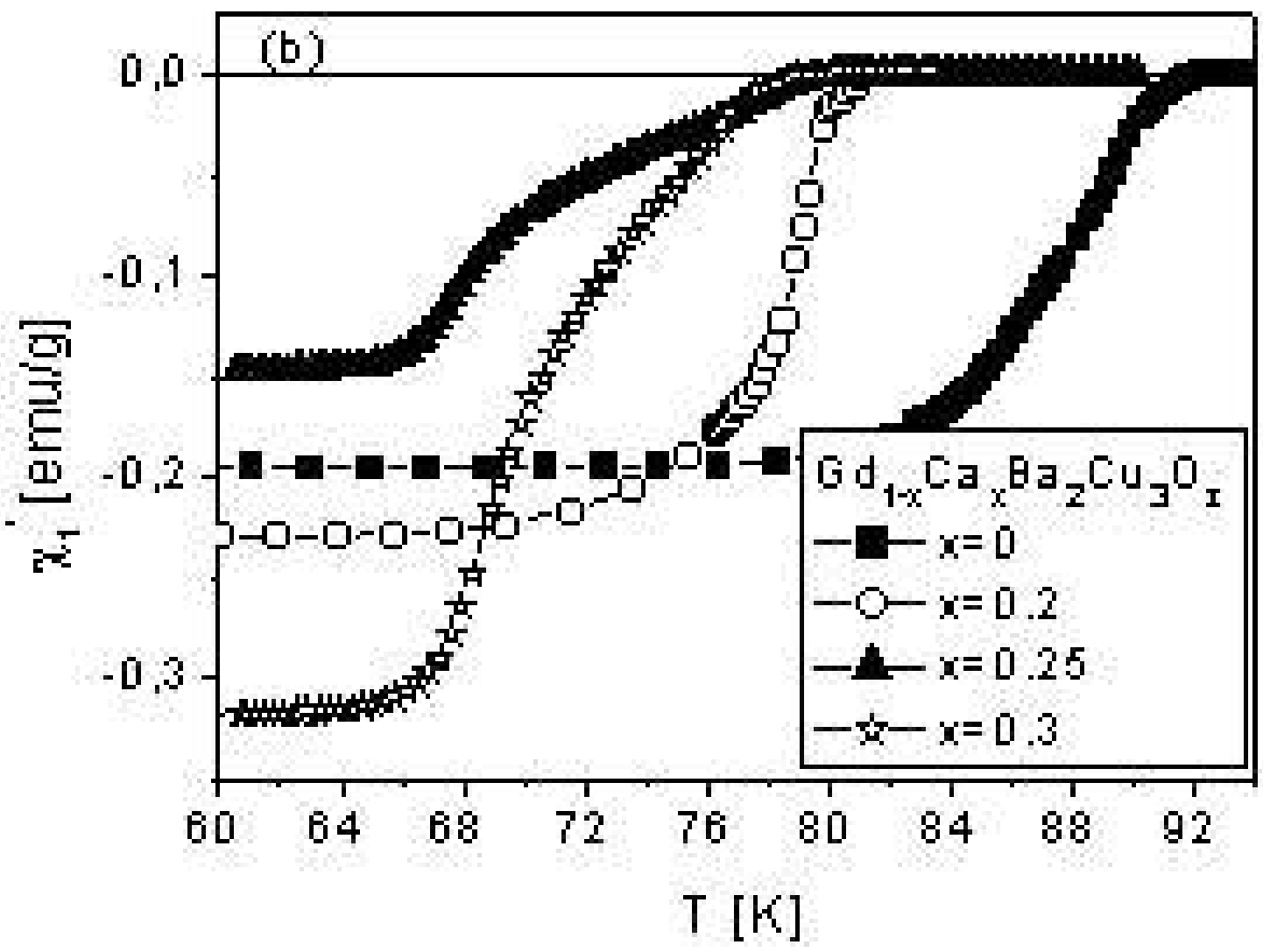
(b)
0,0
-0,1
-0,2
-0,3
χ'1 [emu/g]
60 64 68 72 76 80 84 88 92
T [K]
Gd1-xCaxBa2Cu3Ox
x= 0
x= 0.2
x= 0.25
x= 0.3

Fig.2c.

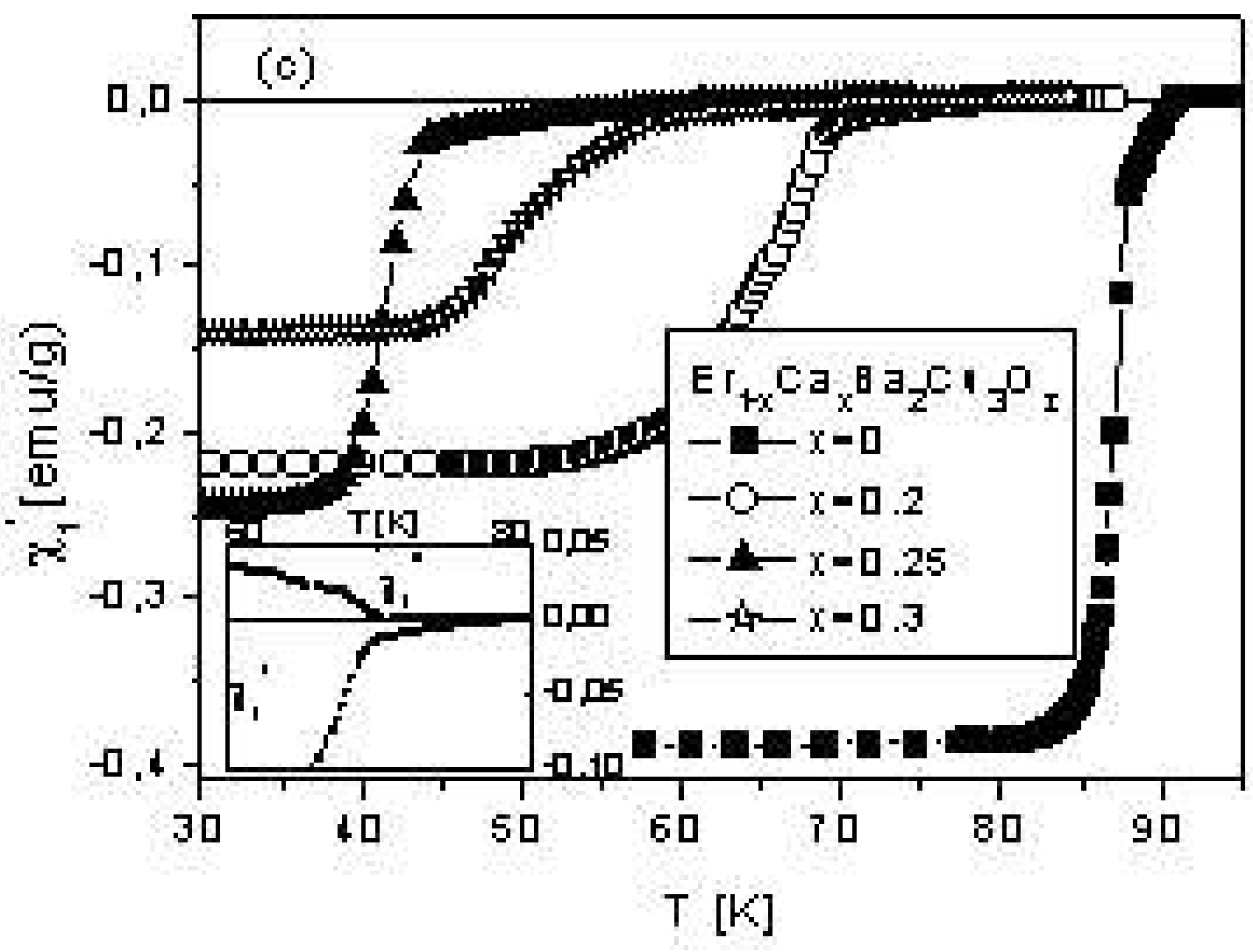

Fig.3.

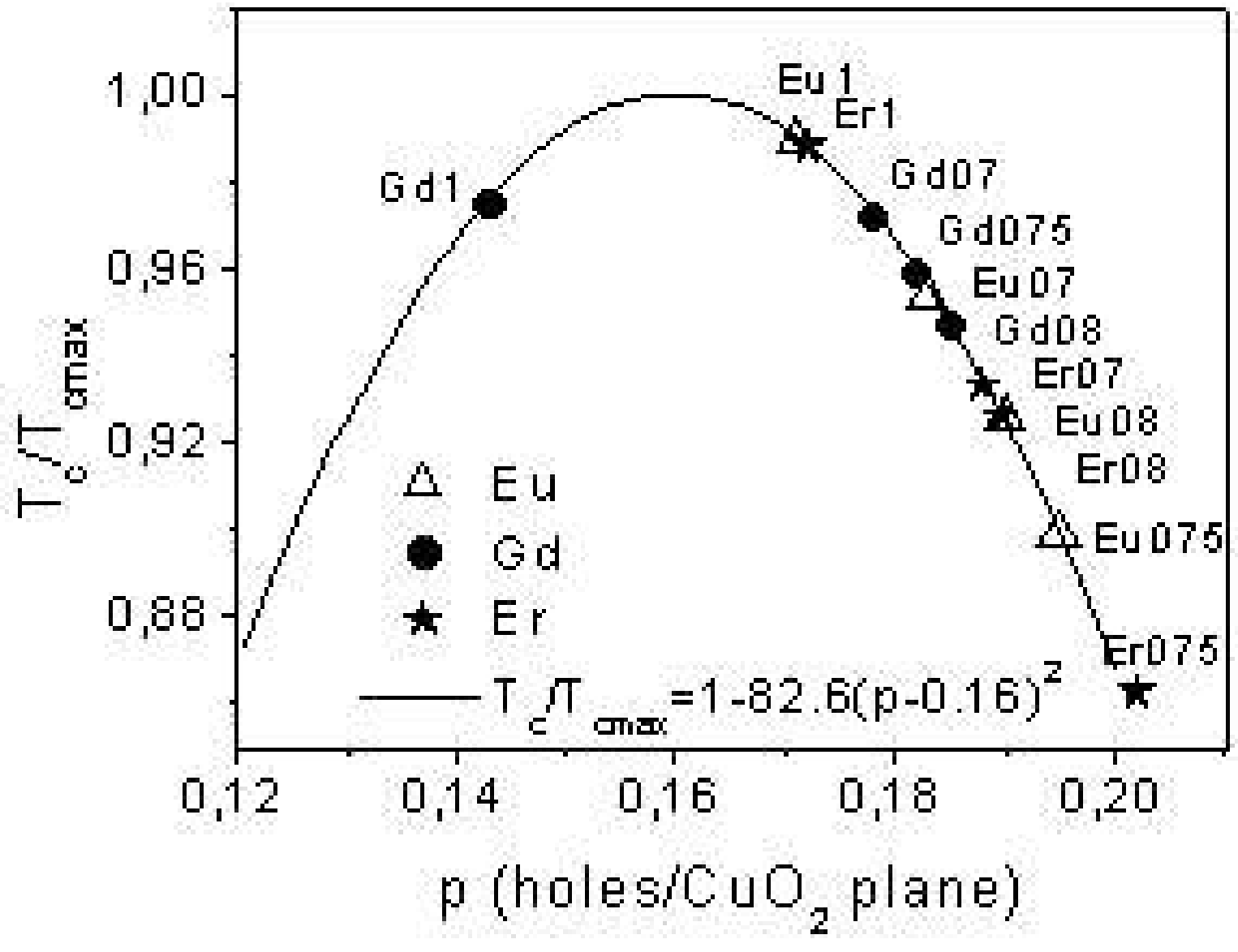

Fig.4.

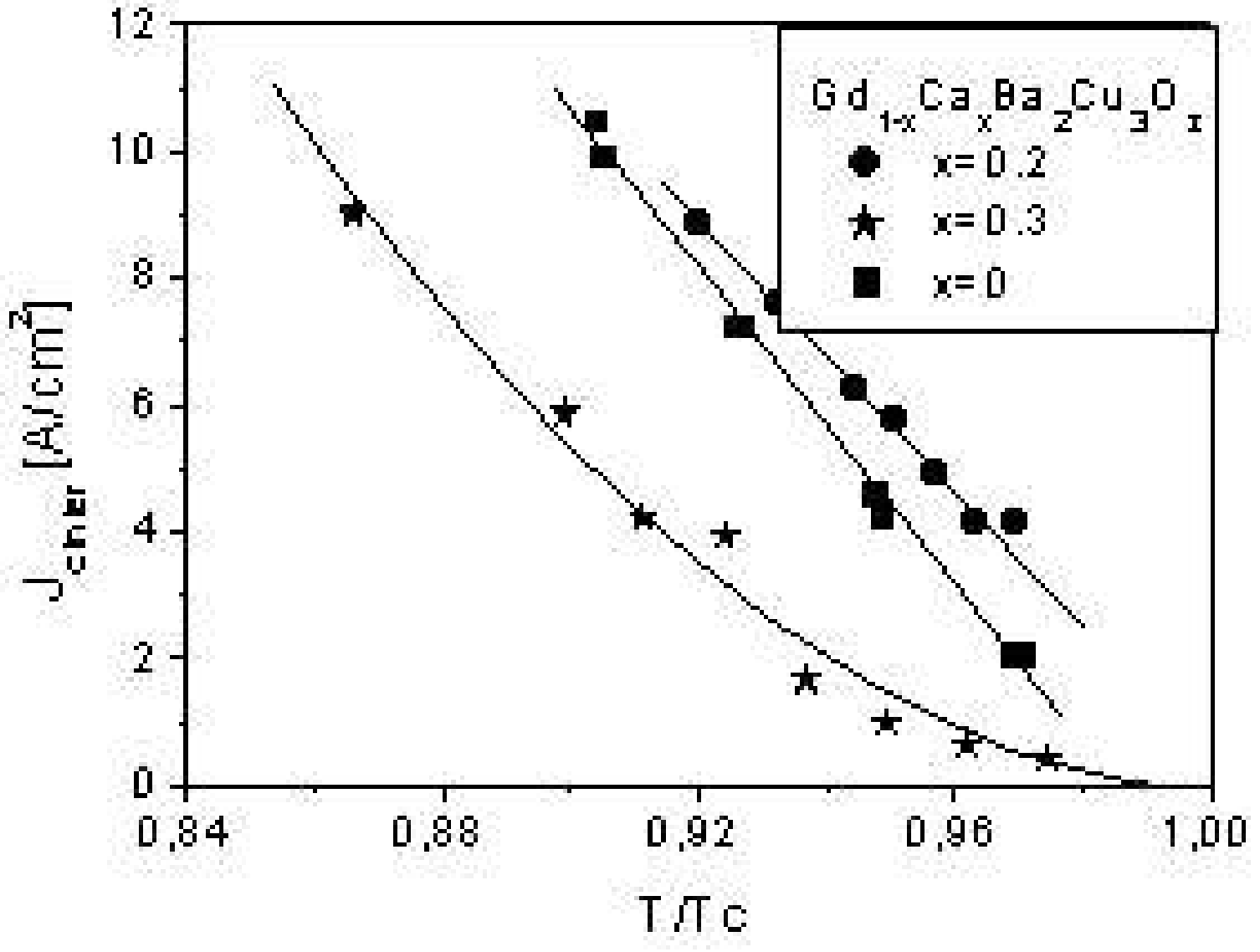

Fig.5a

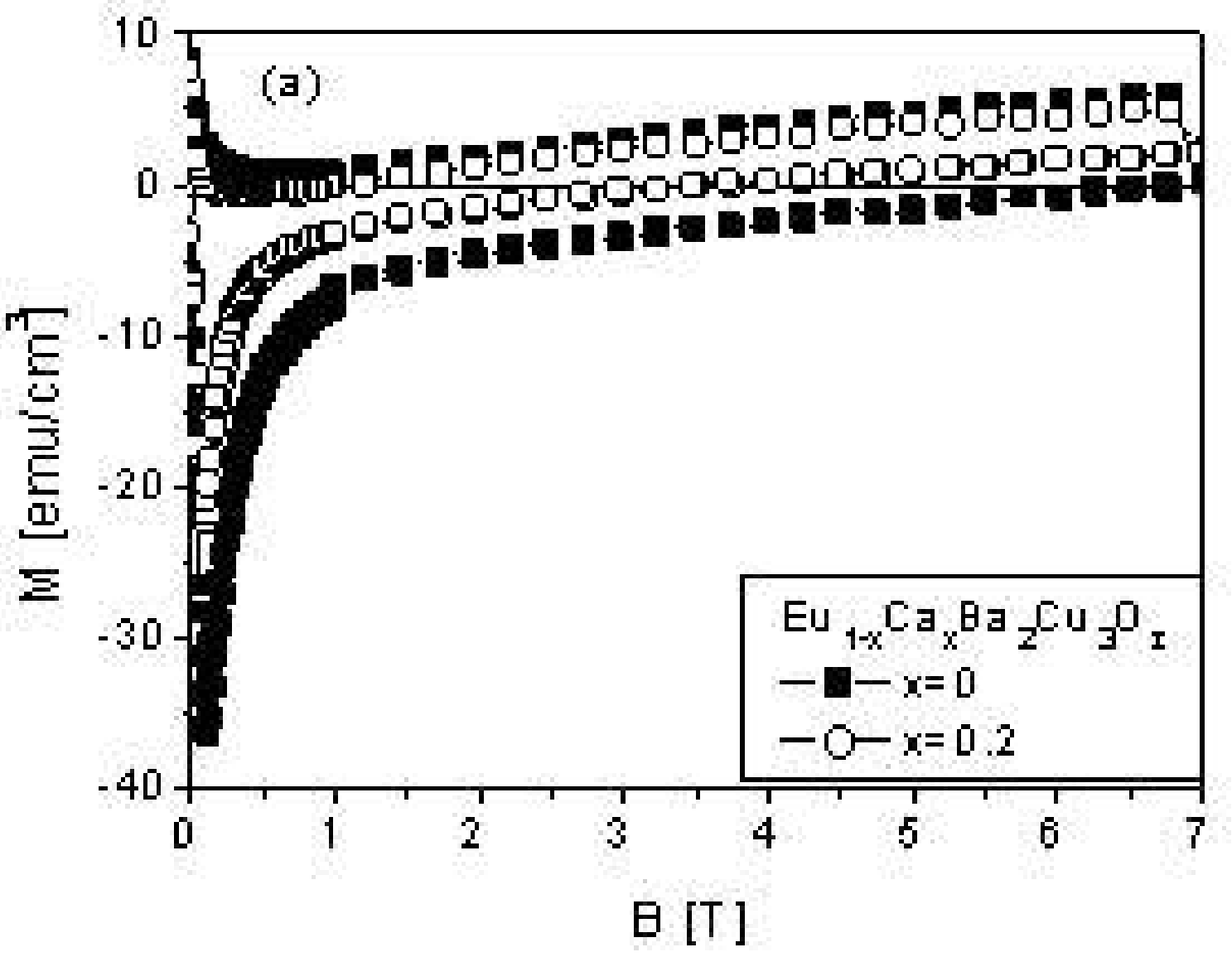
(a)
M [emu/cm³]
B [T]
Eu 1-x Ca x Ba 2 Cu 3 O z
x= 0
x= 0.2
10
0
-10
-20
-30
-40
0
1
2
3
4
5
6
7

Fig.5b

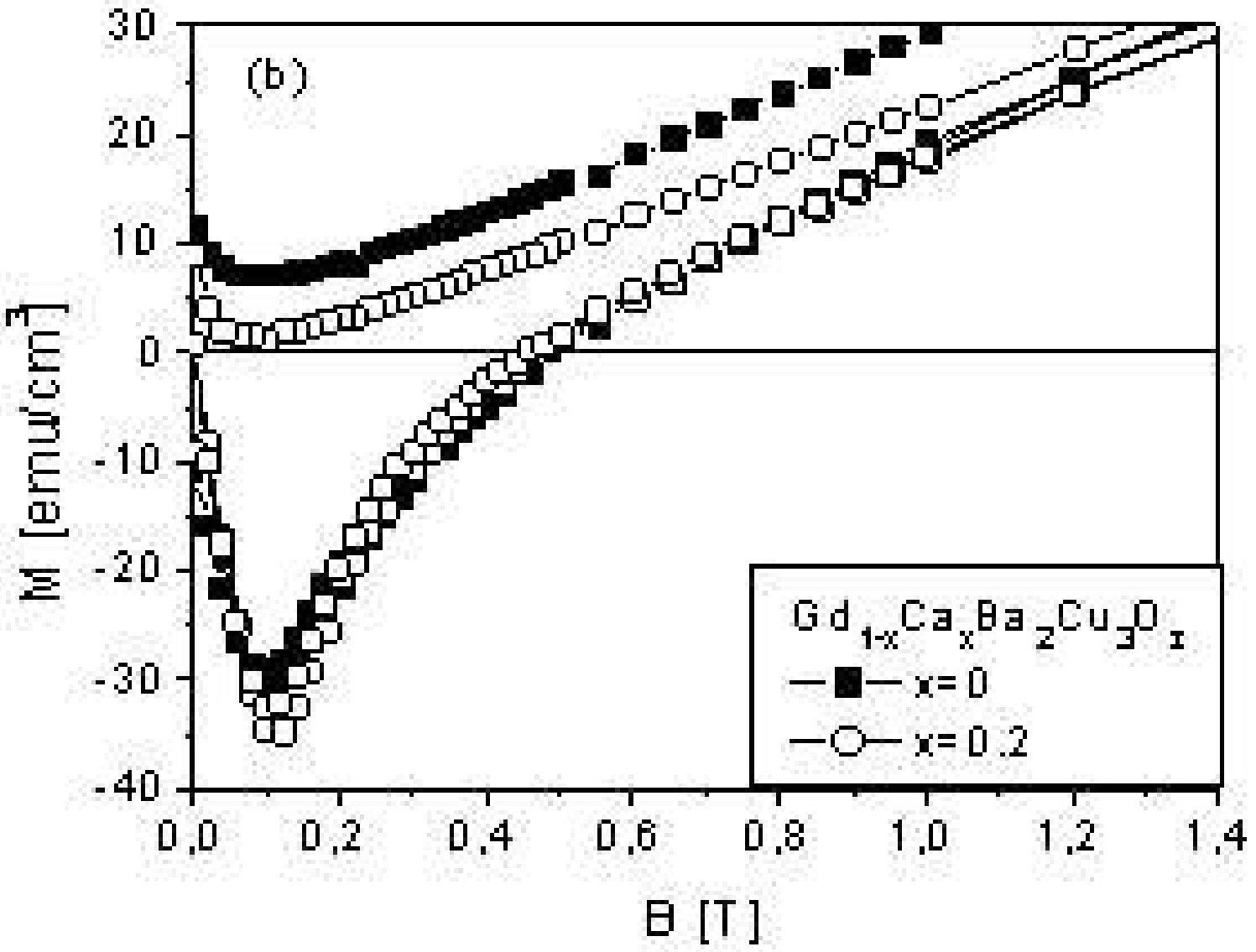
(b)
M [emu/cm³]
B [T]
Gd 1-x Ca x Ba 2 Cu 3 O z
x= 0
x= 0.2
30
20
10
0
-10
-20
-30
-40
0,0
0,2
0,4
0,6
0,8
1,0
1,2
1,4

Fig.5c

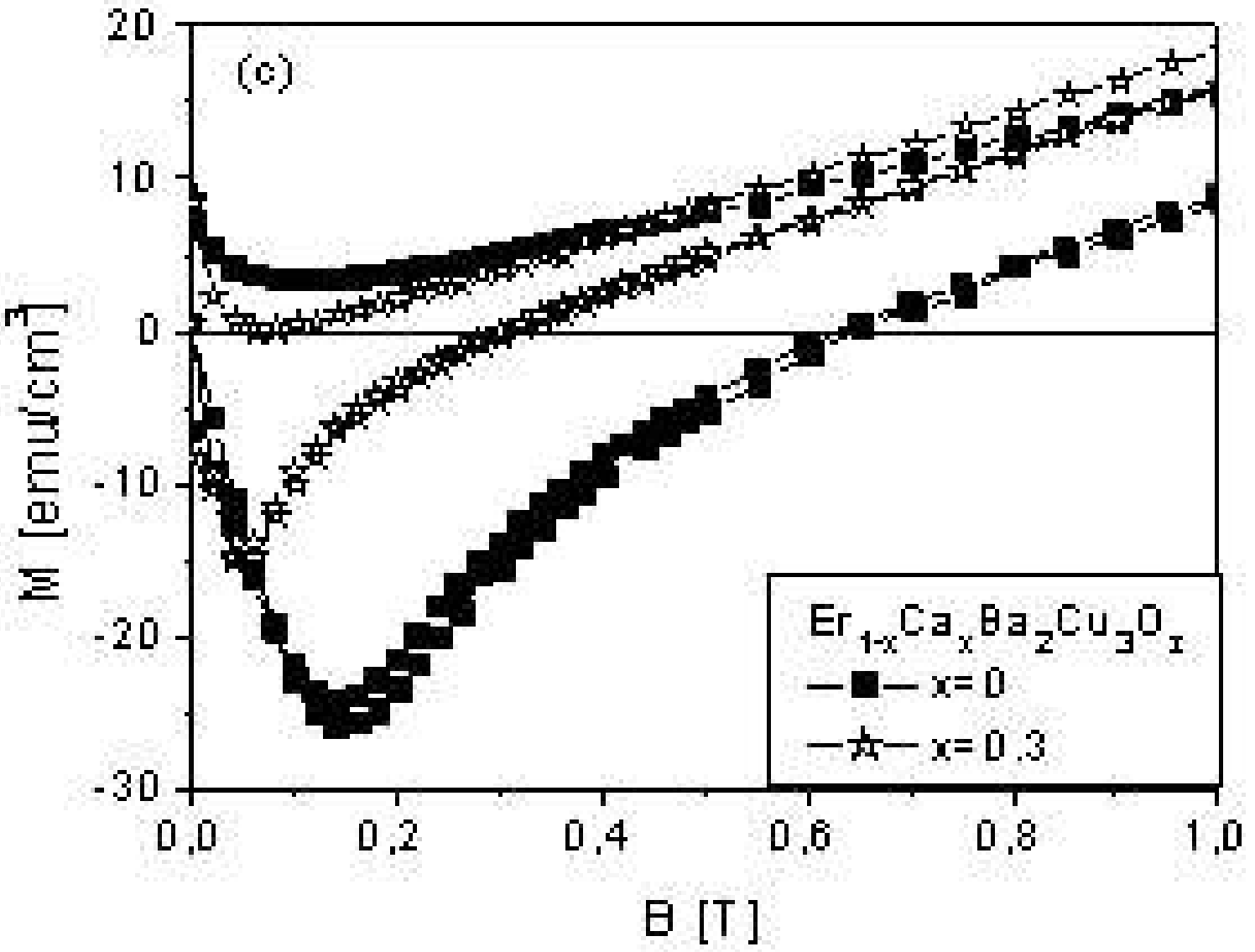

Fig.6.

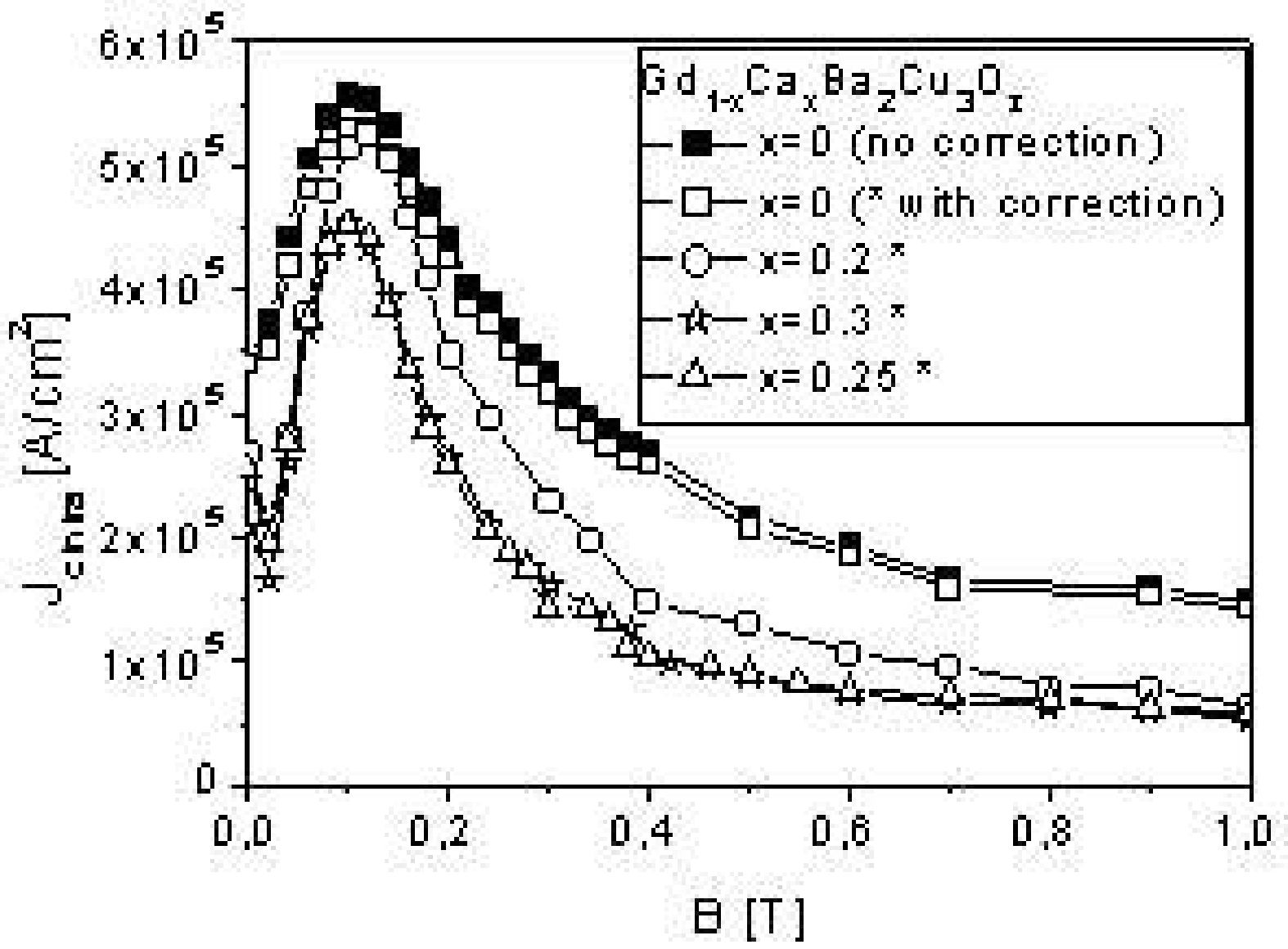

Fig.7.

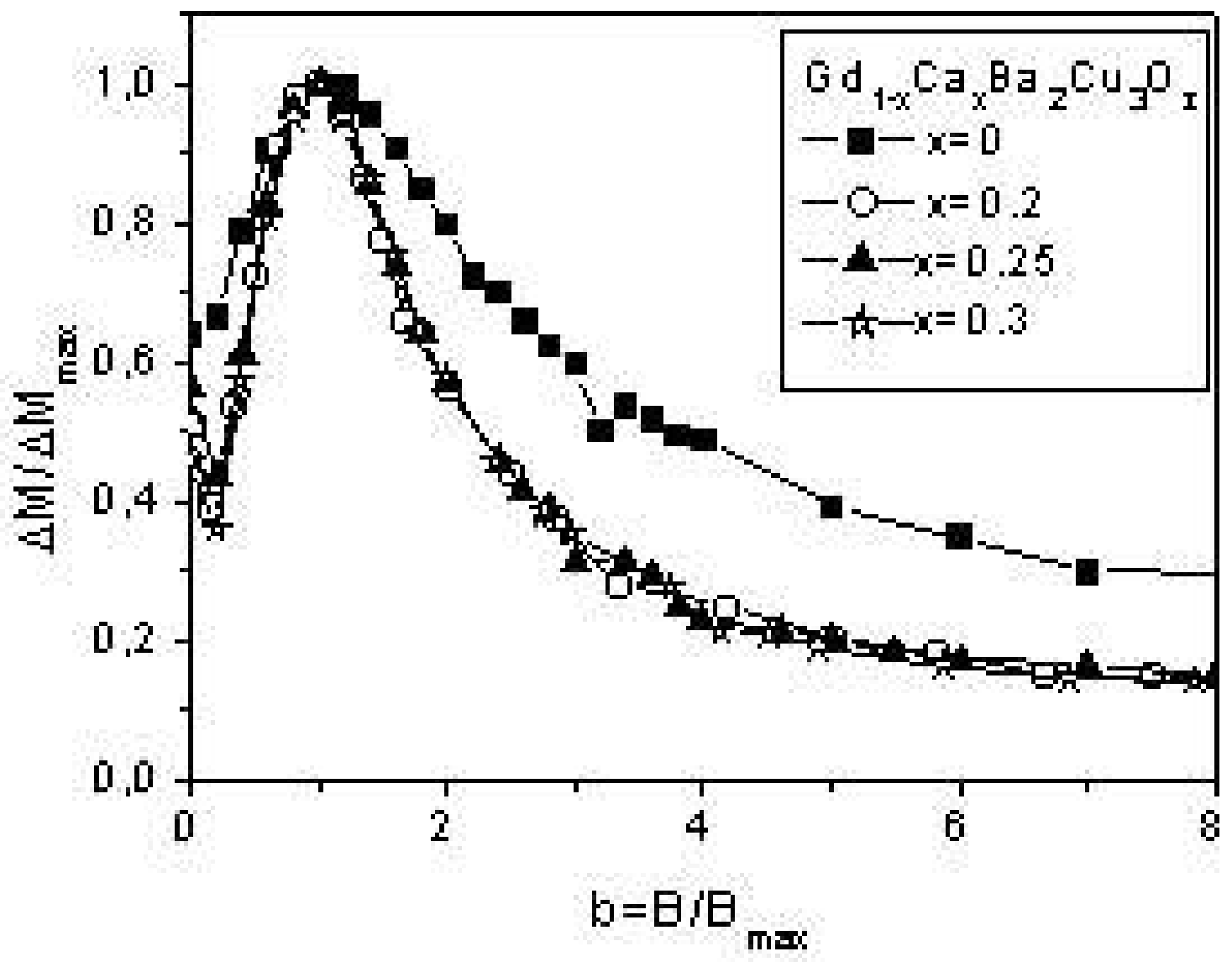

Table I. Lattice parameters *a*, *b* and *c*, orthorhombic splitting (*b*- *a*), $T_{c\ onset}$, $T_{c\ max}$ and hole concentration *p* determined from the ratio $T_c/T_{c\ max}$ for $R_{1-x}Ca_xBa_2Cu_3O_z$ samples.

| Sample | Sym-bol | a (A) | b (A) | c (A) | b- a (A) | $T_{onset}$ (K) | $T_{cmax}$ (K) | p |
|---|---|---|---|---|---|---|---|---|
| $Eu_1Ba_2Cu_3O_z$ | Eu1 | 3.8533 | 3.9060 | 11.7181 | 0.0527 | 93.0 | 94.0 | 0.171 |
| $Eu_{0.8}Ca_{0.2}Ba_2Cu_3O_z$ | Eu08 | 3.8485 | 3.8927 | 11.7099 | 0.0442 | 80.1 | 87.0 | 0.190 |
| $Eu_{0.75}Ca_{0.25}Ba_2Cu_3O_z$ | Eu075 | 3.8549 | 3.8934 | 11.7205 | 0.0385 | 79.7 | 85.25* | 0.195 |
| $Eu_{0.7}Ca_{0.3}Ba_2Cu_3O_z$ | Eu07 | 3.8549 | 3.8802 | 11.7199 | 0.0253 | 79.7 | 83.5* | 0.183 |
| $Gd_1Ba_2Cu_3O_z$ | Gd1 | 3.8432 | 3.9060 | 11.7181 | 0.0628 | 91.7 | 94.0 | 0.143 |
| $Gd_{0.8}Ca_{0.2}Ba_2Cu_3O_z$ | Gd08 | 3.8475 | 3.8996 | 11.7221 | 0.0521 | 81.0 | 85.5 | 0.185 |
| $Gd_{0.75}Ca_{0.25}Ba_2Cu_3O_z$ | Gd075 | 3.8488 | 3.8954 | 11.7418 | 0.0466 | 80.0 | 83.37* | 0.182 |
| $Gd_{0.7}Ca_{0.3}Ba_2Cu_3O_z$ | Gd03 | 3.8428 | 3.8933 | 11.7497 | 0.0505 | 79.0 | 81.25* | 0.178 |
| $Er_1Ba_2Cu_3O_z$ | Er1 | 3.8164 | 3.8868 | 11.6605 | 0.0704 | 90.9 | 92.0 | 0.172 |
| $Er_{0.8}Ca_{0.2}Ba_2Cu_3O_z$ | Er08 | 3.8216 | 3.8817 | 11.6985 | 0.0601 | 78.5 | 84.7 | 0.189 |
| $Er_{0.75}Ca_{0.25}Ba_2Cu_3O_z$ | Er075 | 3.8225 | 3.8762 | 11.6989 | 0.0537 | 71.6 | 82.87* | 0.202 |
| $Er_{0.7}Ca_{0.3}Ba_2Cu_3O_z$ | Er07 | 3.8230 | 3.8752 | 11.6924 | 0.0522 | 75.7 | 81.0* | 0.188 |